\documentclass[12pt]{article}

\usepackage[margin=1in]{geometry}
\usepackage[numbers,sort&compress]{natbib}
\usepackage{graphicx}
\usepackage{amsmath}
\usepackage{booktabs}
\usepackage{xcolor}
\usepackage{multirow}
\usepackage{algorithm}
\usepackage{algpseudocode}
\usepackage{url}
\usepackage{enumerate}
\usepackage{tabularx}
\usepackage{microtype}
\usepackage[colorlinks=true,linkcolor=blue,citecolor=blue,urlcolor=blue]{hyperref}

\newcommand{\MCdC}{MCdC}
\newcommand{\FIFO}{FIFO}
\newcommand{\WRR}{WRR}
\newcommand{\EDAS}{EDAS}
\newcommand{\ARAS}{ARAS}
\newcommand{\MABAC}{MABAC}
\newcommand{\SDK}{SDK}

\title{\textbf{CROWDio}: A Practical Mobile Crowd Computing Framework
  with Developer-Oriented Design, Adaptive Scheduling,
  and Fault Resilience}
\author{
  Lakshani Manamperi$^{*}$ \and
  Disumi Pathirana$^{*}$ \and
  Nipun Premarathna$^{*}$ \and
  Thiwanka Pathirana$^{*}$ \and
  Kutila Gunasekara$^{\dagger}$\\[0.5em]
  \small$^{*}$Undergraduate Student \quad $^{\dagger}$Supervisor\\
  \small Department of Computer Science \& Engineering\\
  \small University of Moratuwa, Sri Lanka\\[0.3em]
  \small\texttt{lakshani.21@cse.mrt.ac.lk} \quad
  \texttt{disumi.21@cse.mrt.ac.lk} \quad
  \texttt{nipunpremarathna.21@cse.mrt.ac.lk}\\
  \small\texttt{thiwanka.21@cse.mrt.ac.lk} \quad
  \texttt{kutila@cse.mrt.ac.lk}
}
\date{}

\begin{document}
\maketitle

\begin{abstract}
Mobile Crowd Computing (\MCdC{}) leverages the idle computational
capacity of consumer smartphones to enable distributed task
processing at scale; however, widespread real-world adoption
remains constrained by the absence of developer-oriented frameworks
capable of transparently managing device heterogeneity, fault
tolerance, and connectivity volatility.
This paper introduces \textbf{CROWDio}, a centralized \MCdC{}
platform comprising three tightly integrated subsystems:
(i)~a declarative \SDK{} that abstracts distributed execution
to a single function annotation, eliminating the need for
explicit parallelism management;
(ii)~a tiered checkpointing mechanism that enables fault-tolerant
task resumption under the memory and execution constraints
inherent to mobile runtimes; and
(iii)~a pluggable multi-criteria scheduling framework driven by
continuous live device telemetry, supporting interchangeable
decision strategies without modification to the dispatch core.
Empirical evaluation across six heterogeneous Android devices
spanning CPU-bound, AI/NLP inference, and data-parallel workloads
demonstrates that capability-aware adaptive scheduling reduces
total execution time by up to 56.9\% relative to na\"ive
round-robin dispatch, while the checkpointing subsystem incurs
a bounded overhead of only 2--3\,s per task regardless of
checkpoint frequency. A system-wide Jain's Fairness Index of
0.889 confirms equitable and stable workload distribution across
heterogeneous worker devices.
\end{abstract}


\section{Introduction}
\label{sec:intro}

Contemporary smartphones routinely operate well below their peak
computational capacity, presenting an opportunity for Mobile Crowd
Computing (\MCdC{}) to aggregate idle resources into a collective
execution platform \citep{pramanik2024mcc}.
Unlike cloud-centric paradigms, \MCdC{} reduces latency and
centralised infrastructure dependency \citep{pramanik2023sustainable}.
Early systems Hyrax \citep{marinelli2009hyrax}, Misco
\citep{dou2010misco}, Honeybee \citep{fernando2013honeybee} proved
the feasibility of smartphone-based distributed computation, but
remained research prototypes without production-grade developer
tooling \citep{kiridana2024mcc}.

The primary barrier to adoption is a \emph{software engineering}
problem: the absence of frameworks that abstract scheduling, fault
tolerance, and device heterogeneity from the developer.
CROWDio addresses this gap with a production-quality \SDK{}
and empirically validated scheduling and checkpointing subsystems.

\textbf{Contributions:}
\begin{enumerate}[(1)]
\item A tiered checkpointing scheme enabling fault-tolerant task
      resumption within constrained mobile runtimes, with bounded
      and predictable overhead.
\item A pluggable multi-criteria decision-making (MCDM) scheduling
      architecture driven by live device telemetry, integrating
      \FIFO{}, \WRR{}, \EDAS{}, \ARAS{}, and \MABAC{} strategies.
\item A modular software architecture following separation-of-concerns
      and pluggable strategy patterns that hides coordination
      complexity from the developer.
\item Empirical evaluation on a six-device heterogeneous cluster
      demonstrating up to 56.9\% scheduling improvement and bounded
      checkpointing overhead.
\end{enumerate}

\section{Related Work}
\label{sec:related}

Hyrax and Misco demonstrated MapReduce on smartphone clusters;
Honeybee introduced work-stealing for dynamic load balancing.
All remained research prototypes without developer \SDK{}s
\citep{pramanik2024mcc,kiridana2024mcc}.
The broader landscape of mobile cloud computing including
computation offloading, resource constraints, and P2P
ad hoc cloud architectures is surveyed comprehensively by
Fernando et al.\ \cite{fernando2013mcc_survey}, who identify the absence of
developer-facing abstractions as a persistent gap.
Similarly, Ray et al.\ \cite{ray2023crowdsensing} survey crowdsensing
and crowdsourcing strategies for smart mobile devices,
distinguishing task-offloading paradigms from crowd-computing
approaches that aggregate peer device capacity.

Volunteer computing systems such as BOINC \citep{anderson2020boinc}
address device heterogeneity, unreliability, and churn at scale,
primarily targeting desktop platforms; their scheduling policies
(resource scheduling, work fetch, job selection) informed the
design of CROWDio's pluggable dispatch architecture.
A recent P2P volunteer computing framework \citep{saleh2022volunteer}
proposes global scheduling based on observed per-device performance,
highlighting the importance of capability-aware dispatch a concern
CROWDio addresses via live telemetry-driven MCDM.

MCDM-based worker selection for heterogeneous \MCdC{} clusters
was introduced by Pramanik et al.\ \cite{pramanik2021mcdm}, with Shannon entropy
weighting \citep{shannon1948} assigned to device capability criteria.
The challenge of scheduling across heterogeneous edge and mobile
resources has also been studied in broader distributed computing
contexts: Stan et al.\ \cite{stan2021scheduling} systematically evaluate
scheduling algorithms in hybrid edge--cloud environments
comprising smartphone and Raspberry Pi nodes, finding that
conventional policies suffer significant performance degradation
on low-capacity edge resources motivating the MCDM approach taken
by CROWDio.
Cloud-oriented distributed frameworks apply clean developer
abstractions \citep{moritz2018ray,zaharia2016spark} but are
unsuited to mobile runtime constraints and volatile connectivity.
CROWDio applies their design philosophy to \MCdC{}:
Table~\ref{tbl:comparison} positions it against prior work.

\begin{table}[h]
\centering
\caption{Comparison of CROWDio with related systems.}
\label{tbl:comparison}
\begin{tabular}{@{}lllll@{}}
\toprule
System & Target & Scheduling & Checkpointing & \SDK{} \\
\midrule
Hyrax/Misco      & Phones  & MapReduce   & None      & No   \\
Honeybee         & Mobile  & Work-steal  & None      & No   \\
BOINC            & Desktop & Static      & App-level & Partial\\
Cloud Frameworks & Cloud   & DAG/Graph   & Lineage   & Yes  \\
\textbf{CROWDio} & \textbf{Mobile} & \textbf{MCDM} & \textbf{Tiered} & \textbf{Yes} \\
\bottomrule
\end{tabular}
\end{table}

\section{System Architecture}
\label{sec:design}

\subsection{Overview}

\begin{figure}[h]
  \centering
  \includegraphics[width=0.85\linewidth]{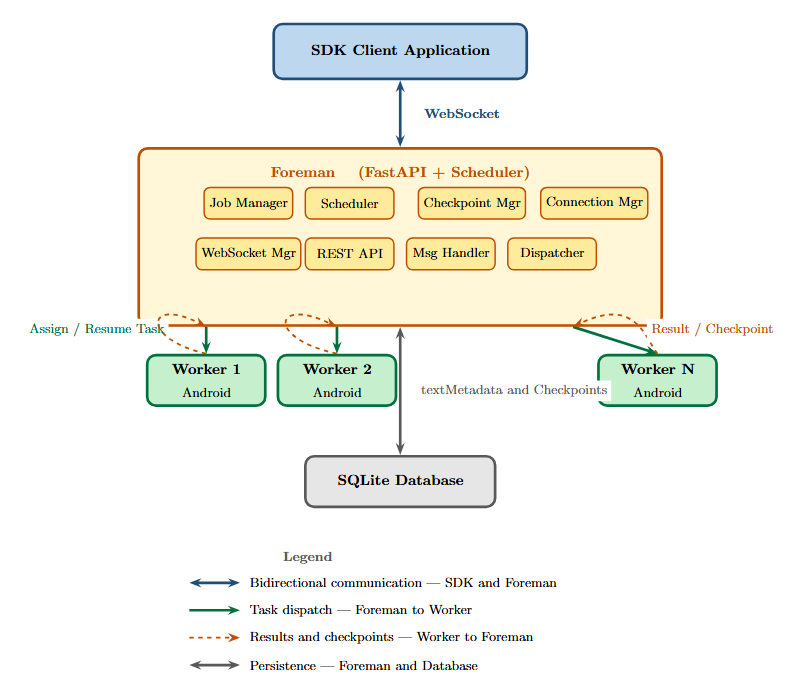}
  \caption{CROWDio high-level architecture. The Coordinator manages
    job decomposition, scheduling, and result aggregation. Workers
    execute tasks on mobile devices; all durable state is persisted
    centrally.}
  \label{fig:architecture}
\end{figure}

CROWDio follows a \emph{Coordinator--Worker} master-worker pattern
across three layers: (1)~the Developer \SDK{}, (2)~the central
Coordinator (control plane), and (3)~distributed mobile Workers
(execution plane). All layers communicate via persistent,
low-latency event-driven connections \citep{rfc6455}.
Figure~\ref{fig:architecture} shows the high-level architecture.

\noindent\textbf{Design principles.}\\
Each Coordinator component carries a single responsibility
(modularity); the scheduling subsystem uses a Strategy pattern
(extensibility); all coordination complexity is hidden from the
developer (abstraction); workers are stateless execution agents
with durable state owned solely by the Coordinator (fault isolation).

\subsection{Developer SDK}

The \SDK{} enables distributed execution through a minimal interface,
requiring only a single function annotation to transform sequential
code into a distributed workload. Parallel operations such as
data-parallel \texttt{map} are dispatched and aggregated
transparently. Fault tolerance checkpointing and automatic
retries, is declared alongside the annotation rather than
implemented by hand, reducing the cognitive overhead of distributed
programming to that of a standard cloud function service.

\subsection{Scheduling}

Worker devices are not interchangeable: a high-end device at low
load is categorically more capable than a low-end device near
saturation. Na\"ive dispatch ignores this, causing systematic
head-of-line blocking. CROWDio instead employs a
\emph{pluggable scheduling architecture} driven by live device
telemetry (CPU utilisation, available memory, battery level,
network latency, and thermal state).

MCDM strategies construct a decision matrix over all connected
workers and observable criteria. Shannon entropy weighting
\citep{shannon1948} assigns criterion weights that reflect
inter-device variability rather than subjective preferences:
\begin{equation}
  w_j = \frac{1 - E_j}{\sum_{k}(1 - E_k)}, \quad
  E_j = -\frac{1}{\ln m}\sum_{i} p_{ij}\ln p_{ij}
  \label{eq:entropy}
\end{equation}
where $p_{ij}=x_{ij}/\sum_i x_{ij}$ and $m$ is the number of
workers. Each pluggable strategy (\EDAS{}, \ARAS{}, \MABAC{}, \WRR{})
ranks workers according to its own decision logic; the highest-ranked
available worker receives the next task. New strategies can be
registered without modifying the core dispatch loop.

\subsection{Fault Tolerance and Checkpointing}

\noindent\textbf{Tiered checkpoint model.}\\
CROWDio uses three checkpoint tiers to balance recovery fidelity
against overhead. \textsc{Base} checkpoints are complete state
snapshots. \textsc{Delta} checkpoints persist only variables that
changed since the last snapshot, minimising transfer and storage
cost. \textsc{Compacted} checkpoints periodically consolidate the
delta chain (default threshold $k=50$), bounding recovery replay
complexity to $O(k)$.

\noindent\textbf{Code instrumentation.}\\
Because standard runtime tracing facilities are unavailable in
constrained mobile execution environments, CROWDio instruments
user-supplied task functions at the source-code representation
level before deployment. Instrumentation: (i)~injects checkpoint
variable recovery at function entry; (ii)~adjusts loop ranges to
skip already-completed iterations on resumption; (iii)~inserts
state-persistence calls after each mutation; and (iv)~wraps
execution in a thread-safe checkpoint state manager. The
transformation is fully transparent to the developer.

\begin{figure}[h]
  \centering
  \includegraphics[width=0.85\linewidth]{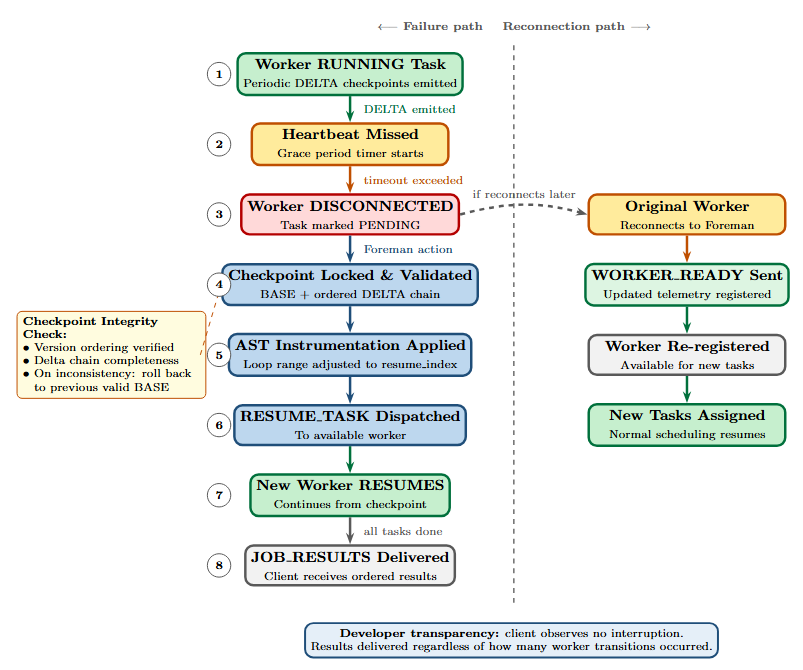}
  \caption{Worker disconnection and graceful reconnection lifecycle.
    Tasks resume on alternative workers from the last checkpoint;
    reconnecting workers re-register with no developer intervention.}
  \label{fig:reconnect}
\end{figure}

\noindent\textbf{Graceful reconnection.}\\
On worker failure, in-progress tasks are transparently resumed on
an alternative worker from the last validated checkpoint.
If the original worker reconnects, it is re-registered immediately;
task ownership is managed atomically by the Coordinator to prevent
duplicate execution. Figure~\ref{fig:reconnect} illustrates the
lifecycle.

\section{Empirical Evaluation}
\label{sec:eval}

\subsection{Experimental Setup}

Six Android devices (Table~\ref{tbl:devices}) were used as workers,
intentionally selected for heterogeneity in clock speed and memory.
Each configuration was repeated ten times; results are reported
as mean $\pm$ SD.

\begin{table}[h]
\centering
\caption{Android worker devices.}
\label{tbl:devices}
\begin{tabular}{@{}llll@{}}
\toprule
Device & Cores & Frequency & RAM \\
\midrule
Samsung Galaxy A34     & 8 & 2.00\,GHz & 7.3\,GB \\
Samsung Galaxy A32     & 8 & 1.80\,GHz & 5.5\,GB \\
Samsung Galaxy A51     & 8 & 1.74\,GHz & 7.4\,GB \\
Motorola E40           & 8 & 1.82\,GHz & 3.4\,GB \\
Samsung Galaxy S6 Lite & 8 & 2.00\,GHz & 3.6\,GB \\
Samsung Galaxy A9+     & 8 & 1.80\,GHz & 3.3\,GB \\
\bottomrule
\end{tabular}
\end{table}

Three workloads were evaluated:
(1)~\textbf{Monte Carlo} estimation of Euler's number $e$ at
1M--100M iterations (CPU-intensive scalability benchmark);
(2)~\textbf{Sentiment Analysis} on a help-desk ticket corpus
\citep{helpdesk} across five workers (AI/NLP inference);
(3)~\textbf{Tile-based Image Processing} on 6{,}862 images
\citep{weatherdata} (data-parallel, coordination-overhead
sensitivity).

\subsection{CROWDio vs.\ Single-Device Execution}

Table~\ref{tbl:crowdio_vs_single} shows that CROWDio's advantage
grows with workload size, achieving $5.1\times$ speedup over the
best single device at 100\,M trials.
Framework coordination overhead is approximately 0.3--0.5\,s per
job negligible at scale. Figure~\ref{fig:comparison} visualises
the widening performance gap.

\begin{table}[h]
\centering
\caption{Single-device vs.\ CROWDio \WRR{} execution time for
  Monte Carlo (mean $\pm$ SD, $n$=10, checkpointing disabled).}
\label{tbl:crowdio_vs_single}
\begin{tabular}{@{}lllll@{}}
\toprule
Trials & Best (s) & Worst (s) & CROWDio (s) & Speedup \\
\midrule
1M   & $2.06\pm0.04$ & $3.12\pm0.05$ & $1.20\pm0.05$ & $\times$1.7 \\
10M  & $19.2\pm0.28$ & $29.7\pm0.43$ & $6.13\pm0.09$ & $\times$3.1 \\
100M & $192\pm2.82$  & $297\pm4.36$  & $37.9\pm0.52$ & $\times$5.1 \\
\bottomrule
\end{tabular}
\end{table}

\begin{figure}[h]
  \centering
  \includegraphics[width=0.85\linewidth]{figs/fig4_comparison_bar.png}
  \caption{Execution time: best device, worst device, and CROWDio
    \WRR{} across six workers. The advantage grows with scale
    ($5.1\times$ at 100\,M trials). Checkpointing disabled.}
  \label{fig:comparison}
\end{figure}

\subsection{Scheduling Algorithm Comparison}

Table~\ref{tbl:scheduling} compares \WRR{} against \FIFO{} for
Monte Carlo. At 100\,M trials, \WRR{} achieves a 56.9\% reduction
(48.9\,s vs.\ 113.4\,s) as \FIFO{} suffers severe head-of-line
blocking on weaker devices. Low standard deviations
($\sigma/\mu < 2\%$) confirm reproducibility across runs.

\begin{table}[h]
\centering
\caption{\FIFO{} vs.\ \WRR{} execution time (mean $\pm$ SD, $n$=10).}
\label{tbl:scheduling}
\begin{tabular}{@{}lllll@{}}
\toprule
Trials & Single (s) & FIFO (s) & WRR (s) & Improvement\\
\midrule
1M   & $2.06\pm0.04$ & $2.0\pm0.06$  & $1.2\pm0.05$  & 40.0\% \\
10M  & $6.13\pm0.09$ & $6.2\pm0.11$  & $5.5\pm0.10$  & 11.3\% \\
100M & $37.9\pm0.52$ & $113\pm2.10$  & $48.9\pm0.89$ & 56.9\% \\
\bottomrule
\end{tabular}
\end{table}

\subsection{Checkpointing Overhead}

Table~\ref{tbl:checkpoint_time} shows a consistent 2--3\,s overhead
regardless of checkpoint interval frequency, attributable to state
serialisation, delta computation, compression, and network
transmission. The 5-second interval provides the best balance
between recovery granularity and performance cost.

\begin{table}[h]
\centering
\caption{Checkpoint frequency vs.\ execution time (1\,M trials,
  mean $\pm$ SD, $n$=10).}
\label{tbl:checkpoint_time}
\begin{tabular}{@{}lll@{}}
\toprule
Mode             & Time (s)      & Overhead   \\
\midrule
Disabled         & $2.06\pm0.04$ & ---        \\
Enabled (5\,s)   & $4.14\pm0.08$ & $+2.08$\,s \\
Enabled (2\,s)   & $4.90\pm0.09$ & $+2.84$\,s \\
Enabled (0.5\,s) & $4.94\pm0.10$ & $+2.88$\,s \\
\bottomrule
\end{tabular}
\end{table}

\noindent\textbf{Coordination-overhead caveat.}\\
Distributed image processing on five workers took substantially
longer than single-device execution due to small per-task batch
sizes. When per-task computation time is comparable to coordination
overhead, distribution yields no benefit analogous to Amdahl's
Law. The ratio of per-task computation to coordination cost must
exceed approximately an order of magnitude for distribution to be
advantageous.

\subsection{Fairness}

Jain's Fairness Index $J = 0.889$ ($\sigma < 0.01$) was consistently
observed across all configurations, confirming stable and
reproducible load distribution despite significant device
heterogeneity.

\section{Discussion and Limitations}
\label{sec:discussion}

CROWDio demonstrates that applying established software engineering
principles to \MCdC{} yields a platform that is simultaneously more
developer-accessible and more maintainable than prior research
prototypes. The single-annotation interface reduces the cognitive
overhead of mobile distributed computing to a level comparable to a
cloud function service. The 56.9\% scheduling improvement confirms
that capability-aware dispatch is essential for heterogeneous mobile
clusters: na\"ive assignment leads to systematic bottlenecking at
scale. The consistent 2--3\,s checkpoint overhead is well justified
in mobile environments where disconnection and battery depletion
are routine events.

\noindent\textbf{Limitations.}\\
The six-device cluster may not capture emergent dynamics at 20--100
devices. The controlled laboratory setting does not replicate
mobility, user interruption, or voluntary participation patterns.
Comparative evaluation of \EDAS{}, \ARAS{}, and \MABAC{} versus
\WRR{} at scale remains future work, as does energy consumption
measurement.

\section{Conclusion}
\label{sec:conclusion}

CROWDio bridges the gap between theoretical \MCdC{} research and
practical deployment by combining a declarative developer \SDK{},
pluggable MCDM scheduling, and tiered checkpointing in a single
mobile-native platform. Empirical evaluation demonstrates linear
scalability, 56.9\% scheduling improvement over na\"ive dispatch,
bounded 2--3\,s checkpointing overhead, and consistent task
distribution fairness all critical for production \MCdC{}
deployment.

Future work will focus on enabling efficient multi-stage deep neural
network inference via pipeline scheduling in mobile crowd computing
environments, extending CROWDio beyond general-purpose computation
toward collaborative, scalable AI inference across mobile devices.

\bibliographystyle{unsrtnat}

\begin{thebibliography}{99}

\bibitem{pramanik2024mcc}
P.K.D. Pramanik, S. Pal, and P. Choudhury.
Mobile crowd computing: potential, architecture, requirements,
challenges, and applications.
\textit{Journal of Supercomputing}, 80(2), 2024.
\url{https://doi.org/10.1007/s11227-023-05545-0}

\bibitem{pramanik2023sustainable}
P.K.D. Pramanik.
Sustainable computing with mobile crowd computing.
Technical Report, NIT Durgapur, 2023.

\bibitem{marinelli2009hyrax}
E.E. Marinelli.
Hyrax: Cloud computing on mobile devices using MapReduce.
Technical Report CMU-CS-09-164, Carnegie Mellon University, 2009.

\bibitem{dou2010misco}
A. Dou et~al.
Misco: A MapReduce framework for mobile systems.
In \textit{Proceedings of PETRA 2010}. ACM, 2010.

\bibitem{fernando2013honeybee}
N. Fernando, S.W. Loke, and W. Rahayu.
Honeybee: A programming framework for mobile crowd computing.
In \textit{MobiQuitous 2012}, Lecture Notes of the Institute for
Computer Sciences, Social Informatics and Telecommunications
Engineering, vol.~120. Springer, 2013.

\bibitem{kiridana2024mcc}
T.B. Kiridana et~al.
Mobile crowd computing with mobile agents and work stealing.
In \textit{Proceedings of MERCon 2024}. IEEE, 2024.

\bibitem{fernando2013mcc_survey}
N. Fernando, S.W. Loke, and W. Rahayu.
Mobile cloud computing: A survey.
\textit{Future Generation Computer Systems}, 29(1):84--106, 2013.
\url{https://doi.org/10.1016/j.future.2012.05.023}

\bibitem{ray2023crowdsensing}
A. Ray, C. Chowdhury, S. Bhattacharya, and S. Roy.
A survey of mobile crowdsensing and crowdsourcing strategies
for smart mobile device users.
\textit{CCF Transactions on Pervasive Computing and Interaction},
5:98--123, 2023.
\url{https://doi.org/10.1007/s42486-022-00110-9}

\bibitem{anderson2020boinc}
D.P. Anderson.
BOINC: A platform for volunteer computing.
\textit{Journal of Grid Computing}, 18(1):99--122, 2020.
\url{https://doi.org/10.1007/s10723-019-09497-9}

\bibitem{saleh2022volunteer}
E. Saleh and C. Shastry.
A new approach for global task scheduling in volunteer computing
systems.
\textit{International Journal of Information Technology}, 2022.
\url{https://doi.org/10.1007/s41870-022-01090-w}

\bibitem{pramanik2021mcdm}
P.K.D. Pramanik et~al.
A comparative analysis of MCDM methods for resource selection in MCC.
\textit{Symmetry}, 13(9):1713, 2021.

\bibitem{shannon1948}
C.E. Shannon.
A mathematical theory of communication.
\textit{Bell System Technical Journal}, 27(3):379--423, 1948.

\bibitem{stan2021scheduling}
R. Stan, C. Copil, D. Moldovan, and H.-L. Truong.
Evaluation of task scheduling algorithms in heterogeneous
computing environments.
\textit{Sensors}, 21(18):6035, 2021.
\url{https://doi.org/10.3390/s21186035}

\bibitem{moritz2018ray}
P. Moritz et~al.
Ray: A distributed framework for emerging AI applications.
In \textit{Proceedings of USENIX OSDI}, pages 561--577, 2018.

\bibitem{zaharia2016spark}
M. Zaharia et~al.
Apache Spark: A unified engine for big data processing.
\textit{Communications of the ACM}, 59(11):56--65, 2016.

\bibitem{rfc6455}
I. Fette and A. Melnikov.
The WebSocket protocol (RFC 6455).
IETF, 2011.
\url{https://doi.org/10.17487/RFC6455}

\bibitem{helpdesk}
M. Abdellatif.
Help desk tickets.
Mendeley Data V2, 2025.
\url{https://doi.org/10.17632/btm76zndnt.2}

\bibitem{weatherdata}
J. Bhathena.
Weather image recognition dataset.
Kaggle, n.d.

\bibitem{edas2017}
M. Keshavarz~Ghorabaee et~al.
Multi-criteria inventory classification using EDAS.
\textit{Informatica}, 26(3):435--451, 2017.

\bibitem{aras2010}
E.K. Zavadskas and Z. Turskis.
A new ARAS method in multicriteria decision-making.
\textit{Technological and Economic Development of Economy},
16(2):159--172, 2010.

\bibitem{mabac2015}
D. Pamucar and G. Cirovic.
The selection of transport resources using MABAC.
\textit{Expert Systems with Applications}, 42(6):3016--3028, 2015.

\end{thebibliography}

\end{document}